\pdfoutput=1

\documentclass[amssymb,amsmath]{iopart} 
\usepackage[english]{babel}
\usepackage{graphics}
\usepackage{graphicx}
\begin{document}

\title{A Transfer Hamiltonian approach in self-consistent field regime for transport in arbitrary quantum dot arrays  } 
\author{S. Illera}
\address{MIND/IN$^2$UB Departament d'Electr\`onica, Universitat de Barcelona, C/Mart\'i i Franqu\`es 1, E-08028 Barcelona, Spain}
\ead{sillera@el.ub.edu}
\author{N. Garcia-Castello}
\address{MIND/IN$^2$UB Departament d'Electr\`onica, Universitat de Barcelona, C/Mart\'i i Franqu\`es 1, E-08028 Barcelona, Spain}
\author{J. D. Prades}
\address{MIND/IN$^2$UB Departament d'Electr\`onica, Universitat de Barcelona, C/Mart\'i i Franqu\`es 1, E-08028 Barcelona, Spain}
\author{A. Cirera}
\address{MIND/IN$^2$UB Departament d'Electr\`onica, Universitat de Barcelona, C/Mart\'i i Franqu\`es 1, E-08028 Barcelona, Spain}
\date{\today}

\begin{abstract}
A transport methodology to study the electron transport between quantum dots arrays based in Transfer Hamiltonian approach is presented. The interactions between the quantum dots and between the quantum dots and the electrodes are introduced by transition rates and capacitive couplings. The effects of the local potential are computed within the self-consistent field regime. The model has been developed and expressed in a matrix form in order to make it extendable to larger systems. Transport through several quantum dot configurations have been studied in order to validate the model. Despite the simplicity of the model, well-known effects are satisfactorily reproduced and explained. The results qualitatively agree with other results obtained using more complex theoretical approaches. 
\end{abstract}

% insert suggested PACS numbers in braces on next line
\maketitle 

\section{Introduction}
Confined structures have been available to the experimentalist for a very long time, the MOS (metal-oxide-semiconductor) transistor is the archetype of a confined two-dimensional system \cite{mos}. Nevertheless, the possibility to enhance this confinement by embedding low-dimensional structures in an insulating matrix has renewed the interest. These structures (quantum dots, wires or layers) can be used in single-electron device \cite{Meir},new memory concepts \cite{Tiwari} and photon or electroluminescent devices \cite{lock}.\\  
Concerning quantum dots (Qds), they are particularly attractive because they possess discrete energy levels and quantum properties similar to natural atoms or molecules. From a fundamental point of view, research has been mostly concentrated on  single quantum dots. These simple systems have been studied using many-body approaches, including non-equilibrium Green's function formalism (NEGFF) \cite{Yey, Meir2}. From a practical point of view, many novel phenomena have been discovered, such as the staircaselike current-voltage (I-V) characteristic \cite{Bar}, Coulomb blockade oscillation \cite{Weis}, negative differential capacitance \cite{Wang}, and the Kondo effect in Qds \cite{vander}. \\
Researchers have recently paid much attention to electron transport through several Qds since multiple Qd provides more Feynman paths for the electron transmission \cite{feyman}. However, up to now the only computation of transport in an extended arbitrary array of Qds was done by Carreras et al. \cite{carr} but no local potential due to self-charge was included. Sun et al.\cite{jap} have also studied the electron transport using NEGFF for different arrangements of Qds, from one to three Qds, without including the potential due to self-charge neither. The inclusion of the self-charge potential using this complex framework is usually impossible for large systems.\\
 In this work, we use non-coherent rate equations (NCRE) \cite{aver, gur} to study the electrical transport in Qds in an extendible, arbitrary, matrix of Qds taking into account self charge effects. In a previous work \cite{mio}, we applied NCRE to obtain analytical solutions for electron transport in simple cases. Using this approach each Qd is treated as a separate system, therefore we can write a NCRE for each dot since these equations describe relationship between the charge inside the Qds and applied bias voltage. The interactions between the Qds, and between the Qds and the electrodes are introduced by transition rates and capacitive couplings. Electron transport and charge densities inside the Qds depend on the tunnel transparency of the barriers limiting each dot. In order to effectively solve the multielectron problem, the effects of the local potential are computed within the self-consistent field (SCF) regime. Moreover, we show how our approach can be easily extended to an arbitrary number of Qds and configurations using a matrix formalism. Therefore, this methodology allows to simulate realistic devices based on large scale Qds arrays. Finally, we compare this methodology with NEGFF, obtaining similar results.   

\section{Theoretical background}
Our system consist of two electrodes (L lead and R lead) coupled to a central transport region. The central region contains several quantum dots, $N$ Qds, distributed inside of an insulator matrix. In order to find the current voltage curve I-V of the total system we use the transfer Hamiltonian formalism\cite{payne, pass}. Using this formalism we can write an expression for the current flowing across two parts of the system. Assuming no inelastic scattering and symmetry in the transmission coefficient \cite{sup1} the net current flux between two parts of the system is 
\begin{equation}
I_{ij}=\frac{4 \pi q}{\hbar}\int T_{ij}(E) \rho_i(E) \rho_j(E) (f_j(E)-f_i(E))dE,
\end{equation}   
where $T_{ij}(E)$ is the transmission probability, $\rho_i(E)$ and $\rho_j(E)$ are the density of states while $f_i(E)$ and $f_j(E)$ are the distribution functions of the different parts of the system. In the equilibrium, the electrochemical potential of the whole system is equal and the particular distribution functions (DFs) are described by the equilibrium Fermi Dirac DF therefore the current between each part of the system is zero. If an external bias voltage (V) is applied, which will drive the system out of the equilibrium, the electrochemical potential of the leads will change by $\mu_L-\mu_R=qV$. From the definition of the total charge $N_i$ inside the $i^{th}$Qd, we can write
\begin{equation}
\label{pop}
N_i=\int \rho_i(E)n_i(E)dE,
\end{equation}
where $n_i(E)$ is an unknown DF and $\rho_i$ is the density of states (DOS) of the $i^{th}$ Qd. For sake of clarity, we only consider one single state with energy level $\epsilon$ in each Qd. In order to take into account the coupling with the surrounding elements we assign a Lorentzian shape DOS centered in $\epsilon$. We can write the evolution charge in time for each Qd as $N_i=\sum_j\int I_{ji}dt$, where the subscript $i$ refers to $i^{th}$Qd and $j$ runs over the other components of the system. Thus, a set of integro-differential equations are obtained for the time charge evolution
\begin{eqnarray}
\label{sist}
\frac{d N_i}{dt}=\frac{4 \pi q}{\hbar}( \int T_{Li}\rho_L\rho_i(f_L-n_i)dE+\int T_{Ri}\rho_R\rho_i(f_R-n_i)dE  \nonumber \\
+\sum_{j\neq i}^{(N-1)} \int T_{ji}\rho_j\rho_i(n_j-n_i)dE) \qquad \forall i=1 \ldots N,
\end{eqnarray} 
where we explicitly write all the current terms: the leads current contributions (first and second term) and the neighbor contribution (the last term). We assume that the DFs in the electrodes ($f_L$ and $f_R$)  are similar to the Fermi Dirac DF using different electrochemical potentials ($\mu_L$ and $\mu_R$). \Eref{sist} can be rewritten for the steady state and assuming no inelastic scattering we can obtain the DF in each Qd for each energy step as a solution of the system of equations 
\begin{eqnarray}
\left(
\begin{array} {ccc}
-T_{L1}\rho_L-T_{R1}\rho_R-\sum_{j\neq 1}^{(N-1)}T_{1j}\rho_j & \ldots & T_{1N}\rho_N  \\ \vdots & \ddots &\vdots \\ T_{1N}\rho_1& \ldots & -T_{LN}\rho_L-T_{RN}\rho_R-\sum_{j\neq N}^{(N-1)}T_{Nj}\rho_j 
\end{array} \right)
\left(
\begin{array}{c} 
n_1 \\\vdots   \\n_N
\end{array}\right) \nonumber \\
 =
\left(
\begin{array}{c} 
-T_{L1}\rho_L f_L-T_{R1}\rho_R f_R \\ \vdots   \\-T_{LN}\rho_L f_L-T_{RN}\rho_R f_R 
\end{array}\right).
\end{eqnarray} 

The effect of the applied voltage to the external electrodes on the electrostatic potential inside each Qd must also be taken into account. The classical solution for the potential at each quantum dot ($V_i$) involves the Poisson equation   
\begin{equation}
\vec{\nabla} \cdot (\varepsilon_r \vec{\nabla} V_i)=-\frac{q\triangle N_i}{\Omega \varepsilon_0},
\end{equation}
where $\varepsilon_r$ is the relative permittivity of the dielectric media, $\varepsilon_0$ is the vacuum permittivity and $\Omega$ is the Qd volume. The general solution for the potential energy $U_i=-qV_i$ in the  $i^{th}$ Qd is \cite{Sup2}
\begin{equation}
\label{pot}
U_i=\sum_{j\neq i}\frac{C_{ij}}{C_{tot,i}}(-qV_j)+\frac{q^2}{C_{tot,i}} \triangle N_i  ,
\end{equation}
where the subscript $j$ runs over all components of the system, $C_{ij}$ is the capacitive coupling between the different components and $C_{tot,i}=\sum_{j, j\neq i}C_{ij}$ is the total capacitive coupling of $i^{th}$Qd. The charge energy constant $U_{0i}=q^2/C_{tot,i}$ is the potential increase as a consequence of the injection of one electron into the Qd and $\triangle N_i$ is the change in the number of electrons, calculated respect to the number of electrons $N_0$ initially in the $i^{th}$Qd.
The effects of local potential on each Qd, which modify the Qd charge and the currents, should be taken into account in the Qd DOS $\rho_i(E) \rightarrow \rho_i(E-U_i)$. In (\ref{pot}) we observe that the local potential depends on the increasing charge density but at the same time the charge depends on the DOS that it is modified by the local potential. These considerations impose a self-consistent solution of (\ref{pop}) and (\ref{pot}).\\

\section{Results and discussion}

In this section, we first show the calculated current voltage curves I-V for different arrangements and we compare their with the results obtained using NEGFF \cite{jap}. We also present the number of electrons $N_i$ accumulated in the $i^{th}$Qd in each configuration. In these cases, analytical expressions for the current are presented as well. Finally, the extension of the model developed in the previous section is presented as a powerful method to study the electron transport in an arbitrary extended array of  Qds.\\
The electrochemical potentials in the two leads are set at $\mu_L=0$ and $\mu_R=-qV$. Electrons flow from the left lead to the right one. For simplicity, we consider that the transmission probability is constant and the same between all the parts of the system. We do not consider direct transmission between the leads. For clarity the DOS of the leads were considered constant in all the energy range. Using this framework, transport without inelastic scattering, the position of the energy levels in the Qds plays an important role therefore the evolution of it with the applied bias voltage define the shape of the I(V) curve. As expected, the I(V) curves exhibit strong dependence with the electrostatic coupling of the different parts of the system. We present expressions for the evolution of the energy level with the applied bias voltage assuming that the elements which are coupled have equal capacitive coupling between them. We set a constant charge energy for all Qds, $U_0=0.25eV$.

\subsection{One single Qd}
\begin{figure}[h!]
\centering{\includegraphics[width=0.5\textwidth]{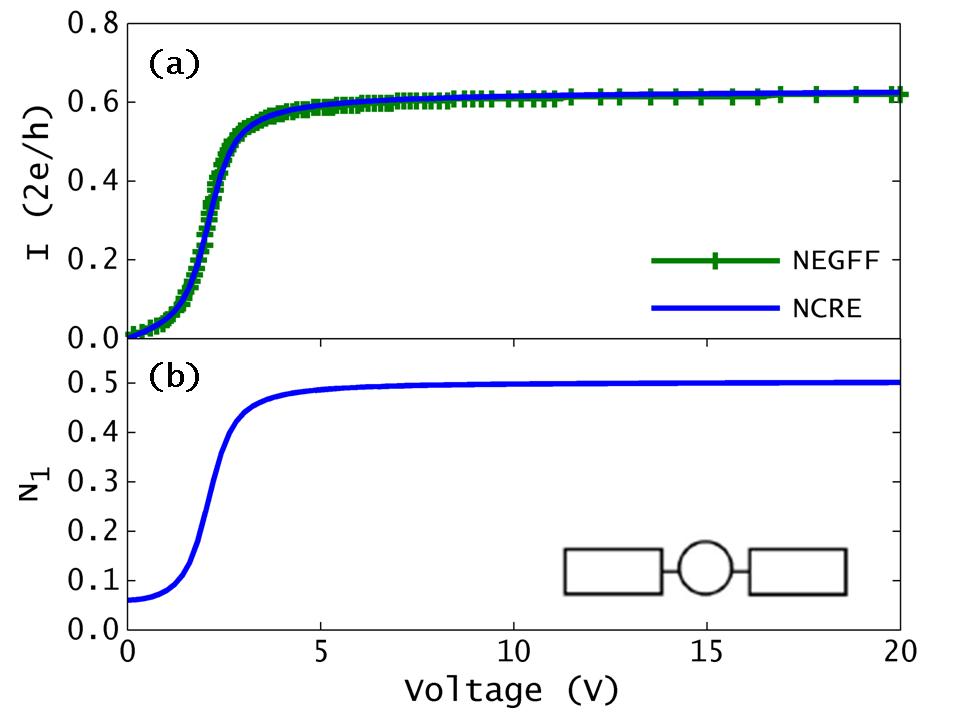}}
\caption{\label{primera}(a) The I-V curve for one single Qd obtained using NCRE. We also show the NEGFF results for the same system, the NEGFF data are taken from Sun et al. \cite{jap}. (b) The electron number in the Qd as a function of the applied bias V. The inset shows the connection geometry. The rectangles represent two leads and the circle represents a Qd.}
\end{figure}
We briefly review electron transport through one Qd. Using (\ref{sist}) and taking into account only lead contributions the current can be written as
\begin{equation}
I=\frac{4 \pi q}{\hbar}\int \frac{T_{R1}T_{L1}\rho_L \rho_1 \rho_R}{T_{L1}\rho_L+T_{R1}\rho_r}(f_L-f_R)dE.
\end{equation}
Fig.~\ref{primera} shows the numerical result of the current I(V). In the calculation we assumed symmetric coupling respect to the leads, $T_{R1}=T_{L1}=0.2$ \footnote{We use similar transmission values than Sun et al.\cite{jap} in order to make possible the qualitative comparison between models.}. The evolution of the energy level with the applied bias voltage is
\begin{equation}
 \epsilon_1(V)=1-V/2+U_0\triangle N_1,
\end{equation}
where the second and third terms are due to the electrostatic effect. As expected the current increases with the bias when the energy of the Qd moves across the left lead; which is $\mu_L=\epsilon_1(V) \rightarrow V \approx 2 $. When $V$ is hight enough, the current saturates to a constant value as $\epsilon_1(V)$ is placed between the two electrochemical potentials of the leads. Fig.~\ref{primera}(b) shows the dependence of the electron number with the applied bias.

\subsection{Two Qds}
We now study the case of two Qds. There are four different connection geometries between Qds and leads. In our calculations we assume symmetric coupling respect to the leads, $T_{R1}=T_{L1}=0.2$, and the Qd coupling $T_{12}=0.2$.

\subsubsection{Parallel case}
\begin{figure}[h!]
\centering{\includegraphics[width=0.5\textwidth]{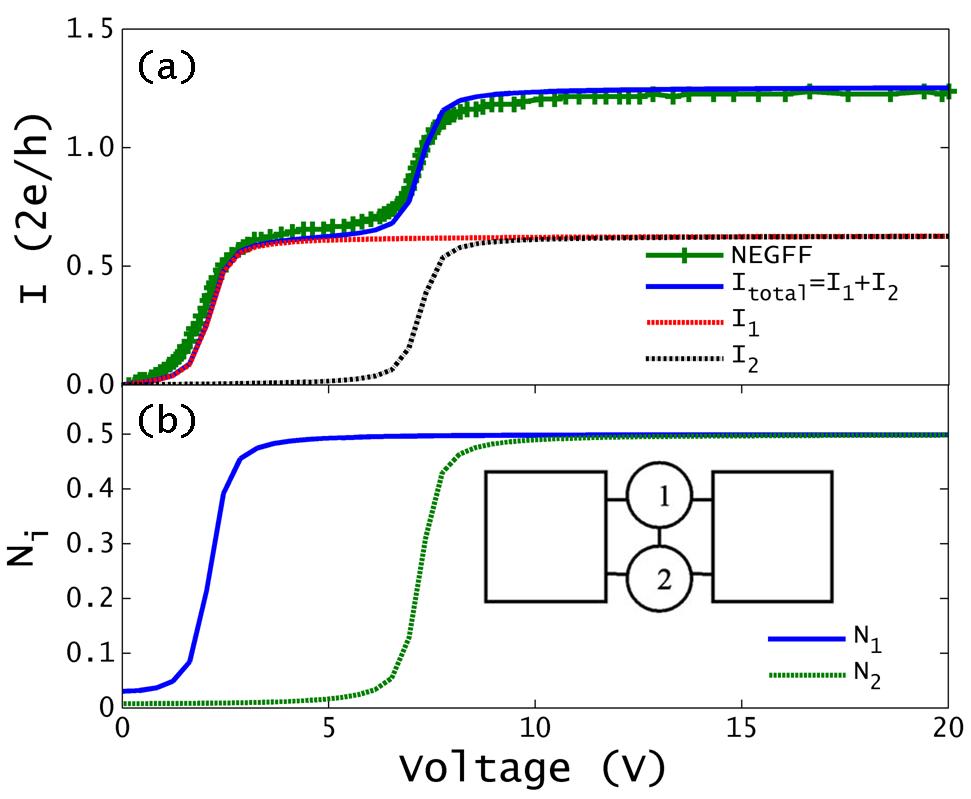}}
\caption{\label{quinta}(a) The total I-V curve and partial I-V curves obtained using NCRE for a parallel configuration. The NEGFF results are taken from Sun et al. \cite{jap}. (b) The electron number in the Qds as a function of the applied bias V.}
\end{figure}
The first configuration of two Qds is the case that they are in parallel. Both Qd are coupled to all elements of the system, the leads and the neighbor Qd. In this configuration the expressions for the current are
\numparts
\begin{eqnarray}
I_1=\frac{4 \pi q}{\hbar}\int \frac{T_{L1}T_{R1}(T_{L1} \rho_L+T_{1R} \rho_R+T_{12}( \rho_1+ \rho_2))\rho_L \rho_R \rho_1}{D_2} \nonumber \\
\times (f_L-f_R)dE\\
I_2=\frac{4 \pi q}{\hbar}\int \frac{T_{L1}T_{R1}(T_{L1} \rho_L+T_{1R} \rho_R+T_{12}( \rho_1+ \rho_2))\rho_L \rho_R \rho_2}{D_2} \nonumber\\
\times(f_L-f_R)dE,
\end{eqnarray}
\endnumparts
where $D_2=(T_{1R} \rho_L+T_{1L}\rho_R)^2+T_{1L}T_{12}\rho_R(\rho_1+\rho_2)+T_{L1} T_{12}\rho_L(\rho_1+\rho_2)$. The position of the energy level of each Qd is
\numparts
\begin{eqnarray}
\epsilon_1(V)=1-qV/3-qV_2/3+U_0\triangle N_1\\
\epsilon_2(V)=3.5-qV/3-qV_1/3+U_0 \triangle N_2.
\end{eqnarray}
\endnumparts
We show the total and partial currents Fig.~\ref{quinta}(a). The I-V curve shows two steps when the energy levels of the Qds are placed between the electrochemical potentials of the leads. This case is equivalent to a single Qd with two energy levels. Fig.~\ref{quinta}(b) shows the electron number $n_i$ with the applied bias voltage. The charge increases until it reach the saturation value.

\begin{figure}[h!]
\centering{\includegraphics[width=0.5\textwidth]{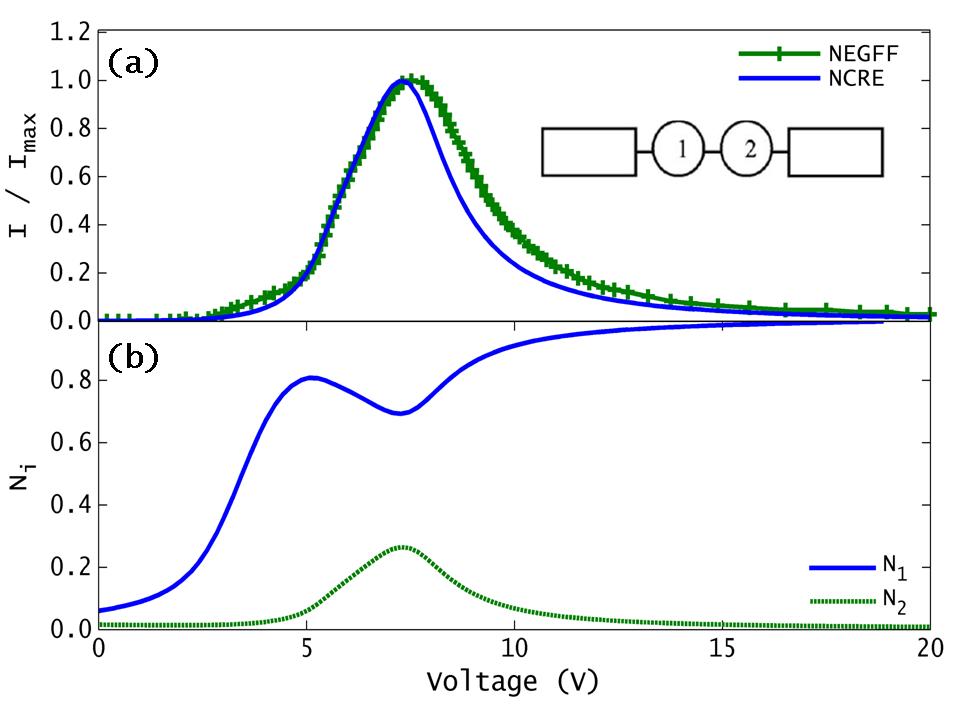}}
\caption{\label{segunda}(a) The I-V curve for two Qds in a serial configuration obtained using NCRE. We also show the NEGFF results for the same system, the NEGFF data are taken from Sun et al. \cite{jap}. The inset shows the connection geometry. (b) The electron number in the Qds as a function of the applied bias V.}
\end{figure}
\subsubsection{Serial case}
The second type of arrangement is the case of two Qds in a serial configuration. The system is shown in the inset of Fig.~\ref{segunda}(a). Each Qd only interacts with one lead and the other Qd. In this case, the expression for the current  is
 \begin{eqnarray}
I=\frac{4 \pi q}{\hbar}\int \frac{T_{L1}T_{12}T_{2R}\rho_L \rho_1 \rho_2 \rho_R}{T_{L1}T_{12}\rho_1 \rho_L+T_{L1}T_{2R}\rho_R \rho_L+T_{12}T_{2R}\rho_2 \rho_R}\nonumber\\
\times  (f_L-f_R)dE
\end{eqnarray} 
and the evolution of the energy level of each Qd with the applied bias voltage is
\numparts
\begin{eqnarray}
\epsilon_1(V)=1-qV_2/2+U_0\triangle N_1\\
\epsilon_2(V)=3.5-qV/2-qV_1/2+U_0\triangle N_2,
\end{eqnarray}
\endnumparts
where we assumed that the Qds are only coupled to each other and to one lead. In order to have current flowing through the system, the energy levels must lie between the electrochemical potentials of the leads and overlapping of the Qd energy levels is necessary. This means that the electrons need available states in each part of the system in order to move from Left lead to Right lead. When the energy levels are equal, $\epsilon_1=\epsilon_2 \rightarrow V \approx 7.5 $, this is a maximum overlapping between Qd DOS, the current is maximum and the system is in a resonance state therefore the channel is open. When the voltage increases further the Qd DOS overlapping decreases. Therefore a negative differential resistance appears \cite{borg}. In Fig.~\ref{segunda}(b) we show the evolution of the charge in each Qd $N_i$ as a function of the applied voltage V. Initially, $N_1$ increases since the channel between the first and second Qd is closed. At the resonant condition, the channel between the Qds opens and some charge stored in the first Qd flows to the second Qd. At higher voltages the channel closes again and $N_1$ stores all the incoming charge, while $N_2$ loses its charge. 

\subsubsection{Other two Qds configurations}
\begin{figure}[h!]
\centering{\includegraphics[width=0.5\textwidth]{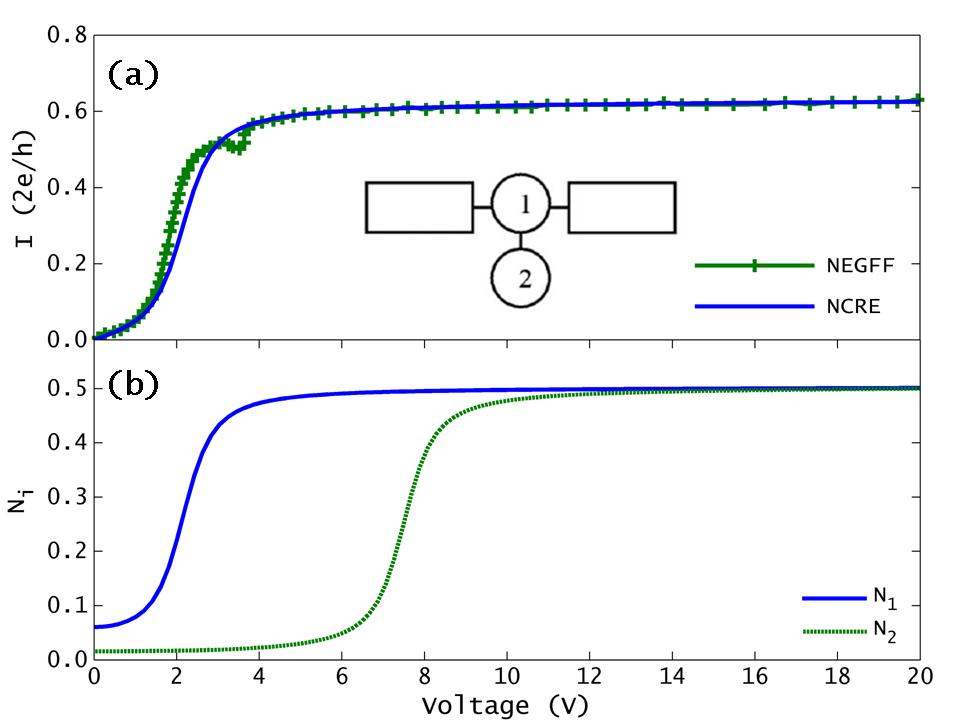}}
\caption{\label{tercera}(a) The I-V curve, for the configuration plotted in the inset, obtained using NCRE. We also show the NEGFF results for the same system, the NEGFF data are taken from Sun et al. \cite{jap}. (b) The electron number in the Qds as a function of the applied bias V.}
\end{figure}
We first examine the case in which one Qd interacts with the two leads and it is also connected to the second Qd, while the second Qd is only connected to the first Qd. The current is
\begin{equation}
\label{I2}
I=\frac{4 \pi q}{\hbar}\int \frac{T_{R1}T_{L1}\rho_L \rho_1 \rho_R}{T_{L1}\rho_L+T_{R1}\rho_r}(f_L-f_R)dE
\end{equation}
and the position of the energy levels are
\numparts
\begin{eqnarray}
\epsilon_1(V)=1-qV/3-qV_2/3+U_0\triangle N_1\\
\epsilon_2(V)=3.5-qV_1+U_0\triangle N_2.
\end{eqnarray}
\endnumparts
The obtained current expression \ref{I2} is the same than the one we obtained for the single Qd case. The DF in the second Qd is the same as in the first Qd therefore the current between the Qds is zero. The results are presented in Fig.~\ref{tercera}.\\
\begin{figure}[h!]
\centering{\includegraphics[width=0.5\textwidth]{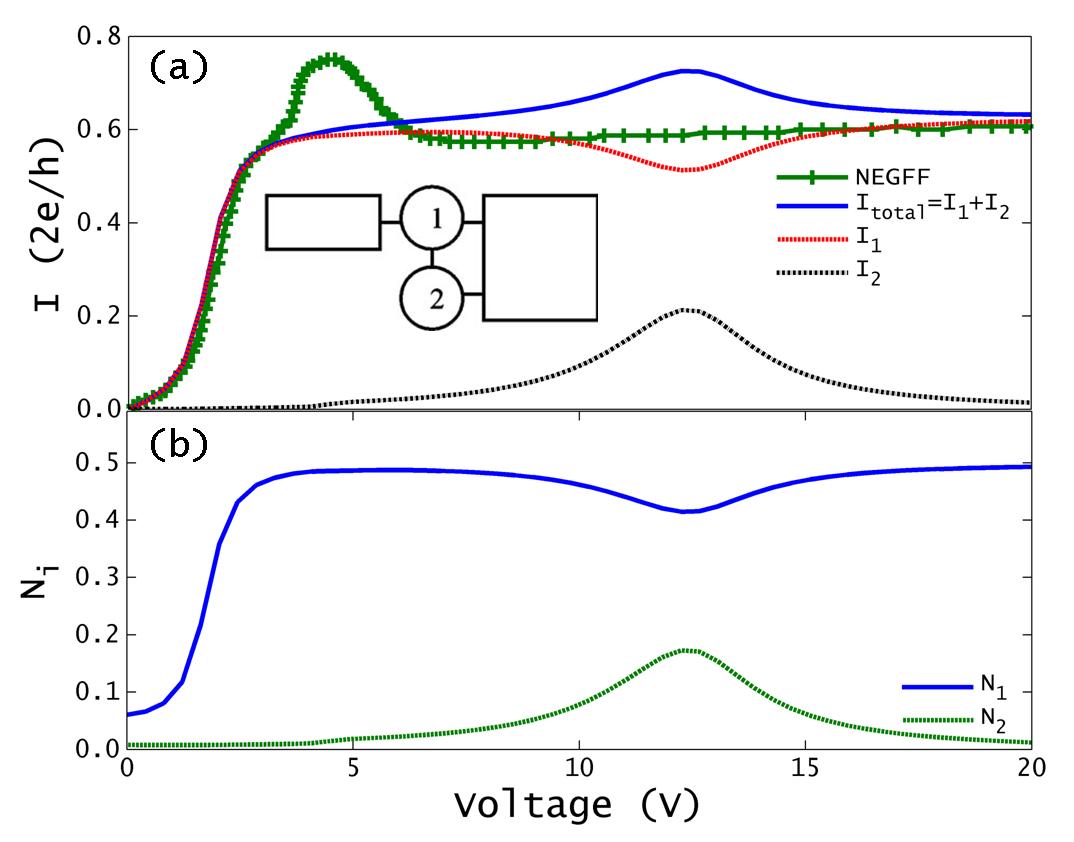}}
\caption{\label{cuarta}(a) The total I-V curve and partial I-V curves obtained using NCRE for the configuration showed in the inset. The NEGFF results are taken from Sun et al. \cite{jap}. (b) The electron number in the Qds as a function of the applied bias V. }
\end{figure}

The second arrangement of Qds is shown in the inset of Fig.~\ref{cuarta}. The expressions for the current are
\numparts
\begin{eqnarray}
I_1=\frac{4 \pi q}{\hbar}\int \frac{T_{1R} \rho_1 \rho_R (T_{R2} \rho_R T_{L1} \rho_L+T_{12} \rho_1 T_{L1}\rho_L)}{D}(f_L-f_R)dE\\
I_2=\frac{4 \pi q}{\hbar}\int \frac{T_{2R}\rho_2 \rho_R T_{12} \rho_1 T_{L1} \rho_L}{D}(f_L-f_R)dE,
\end{eqnarray}
\endnumparts
where $D=T_{2R}\rho_RT_{L1}\rho_L+T_{R1}T_{R2}\rho_R^2+T_{R2}\rho_R T_{12}\rho_2+T_{12}\rho_1T_{L1}\rho_L+T_{R1}\rho_R T_{12} \rho_1$ and the total current is $I=I_1+I_2$. The energy level position is
\numparts
\begin{eqnarray}
\epsilon_1(V)=1-qV/3-qV_2/3+U_0\triangle N_1\\
\epsilon_2(V)=3.5-qV_1/2+U_0\triangle N_2.
\end{eqnarray}
\endnumparts
In this case we show the total and partial currents. The I-V partial current shows an interesting behavior. The current  through the first Qd is similar to the single one Qd configuration but the current through the second reminds the slope of a resonant state. This fact can be easily understood in the following way: if the channel between the two Qds is closed the current only flows through the first Qd. When the Qd1-Qd2 channel is opened the Qd2 also conducts. In the same case as before, when the voltage increases the overlapping decreases and the Qd2 current decreases. 

\subsection{Three Qds}
\begin{figure}[h!]
\centering{\includegraphics[width=1\textwidth]{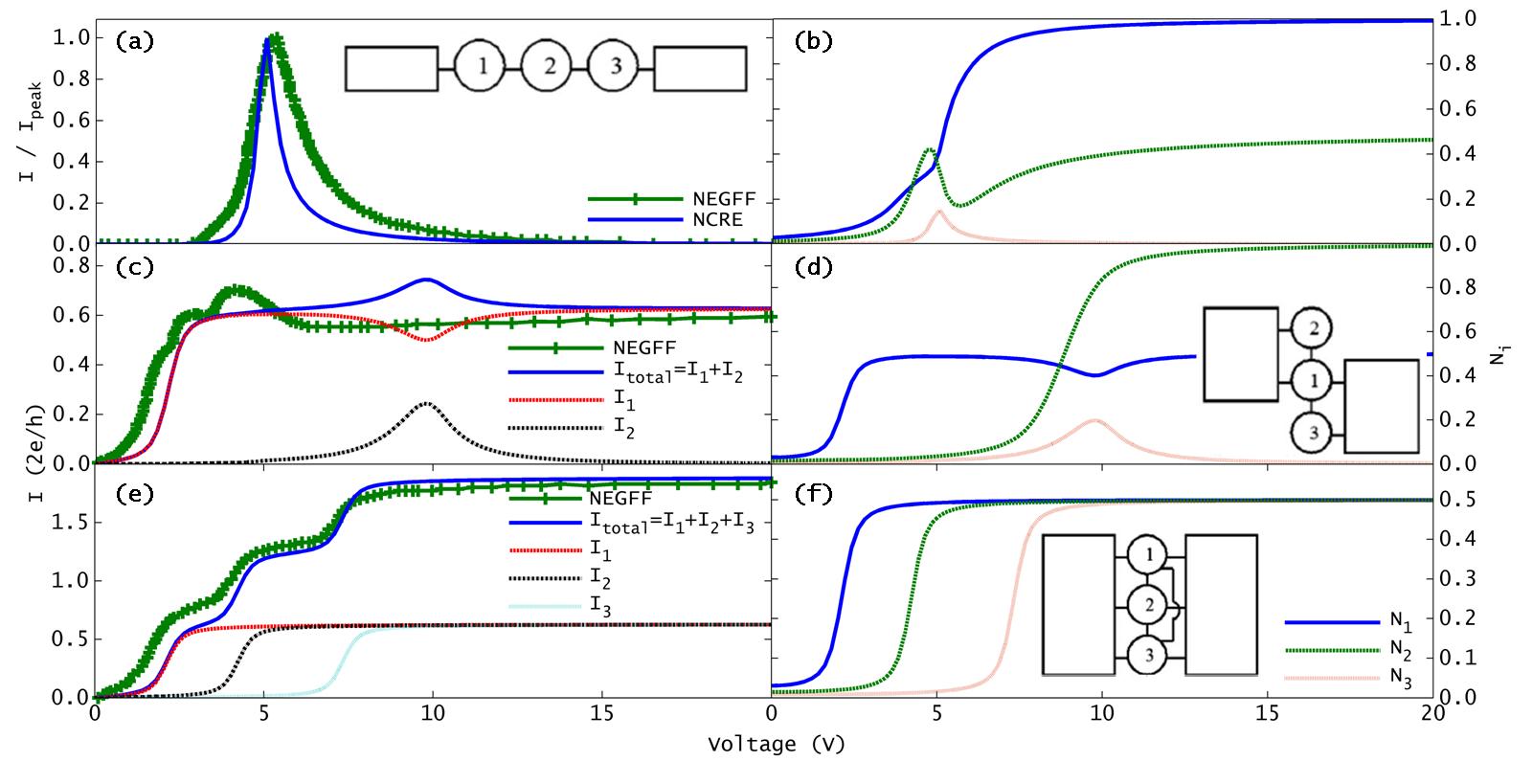}}
\caption{\label{sexta}[(a),(c) and (e)] The I-V curves and the electron number [(b), (d) and (f)] respectively for three Qds with different configuration. The insets show the connection geometry. The NEGFF results are taken from Sun et al. \cite{jap}.}
\end{figure}
The methodology developed in the first section can be easily extended into more complicated systems. Here, we present the results for some configurations based in three Qds. The analytical expressions for the current are too large to write here but in Fig.~\ref{sexta} [(a),(c),(e)] we show the I-V curves and the charge in each Qd Fig.~\ref{sexta}[(b),(d),(f)]. As we have shown before the position of the energy levels plays an important role in the I-V and N-V curves, using \ref{pot} we can write the position of the each energy level as a function of the applied bias voltage
\numparts
\begin{eqnarray}
\epsilon_1(V)=1-\sum_{j}\frac{C_{1j}}{C_{total1}}V_j+U_0\triangle N_1\\
\epsilon_2(V)=2-\sum_{j}\frac{C_{2j}}{C_{total2}}V_j+U_0 \triangle N_2\\
\epsilon_3(V)=3.5-\sum_{j}\frac{C_{3j}}{C_{total3}}V_j+U_0\triangle N_3,
\end{eqnarray}
\endnumparts
where the subscript $j$ runs over all connected elements of the system. The Qd-lead coupling and the interdot coupling are set equal $T_{ij}=0.2$.  In the insets of the Fig.~\ref{sexta} we show the scheme of the system under study.

\subsection{Large Qds arrangement}

\begin{figure}[h!]
\centering{\includegraphics[width=0.5\textwidth]{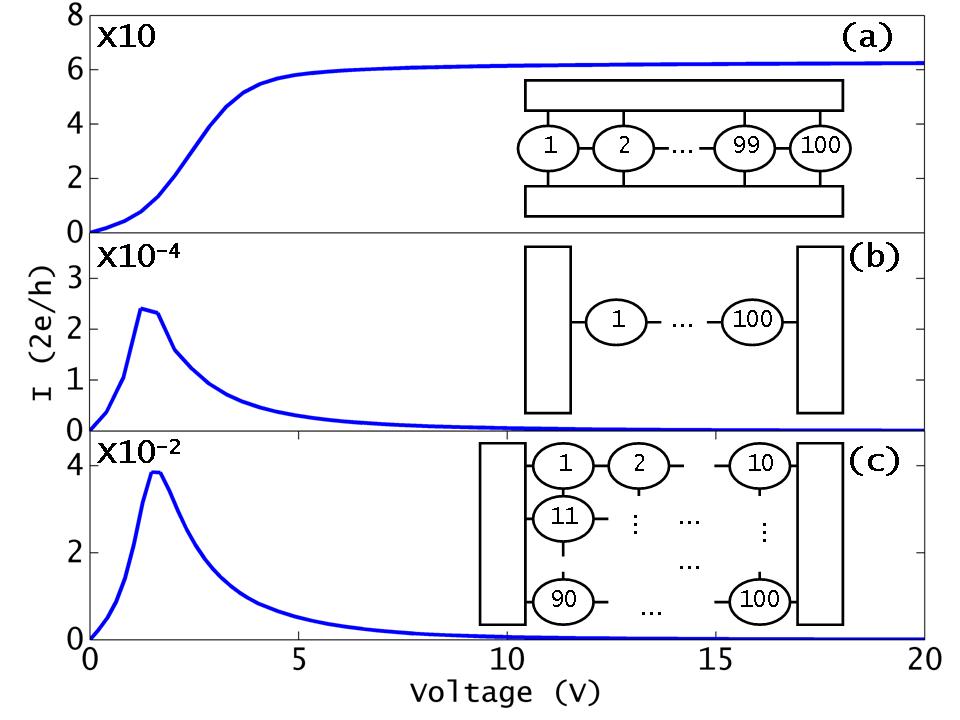}}
\caption{\label{septima}The I-V curves for the larger systems: (a) 100 Qds in parallel configuration, (b) 100 Qds in serial configuration and (c) 100 Qds in an array disposition $10\times 10$.}
\end{figure}

To conclude we present the results for larger systems that they are close to the experimental measurements. The systems are formed by 100 Qds placed in a parallel configuration, serial configuration and in an array geometry ($10 \times 10$). The total I-V curves and the geometries are presented in Fig.~\ref{septima}. The Qd-lead coupling and the interdot coupling are set equal $T_{ij}=0.2$. The capacitance between the linked elements are also equal. In order to represent an experimental system we considerer that the value of the energy level of each dot follows a normal distribution with mean value $1$eV and deviation $0.2$eV. This fact represents the usual distribution size that appears in the experiments. The relationship between the Qd radius and the energy level position is a well known effect and it is related to the quantum confinement of the electrons \cite{delerue}.\\
The I-V curves show an interesting behavior. First, in the parallel case, Fig.~\ref{septima}(a), the I-V curve shows a staircaselike structure and saturates to a constant value at high bias. As we have seen before in the parallel configuration each Qd acts as an independent channel therefore the total current is the sum of all partial currents. As expected, the saturation current is 100 times the saturation current of a single dot. \\
For the serial configuration, Fig.~\ref{septima}(b), we obtain a current peak as we expect due to the resonant state is necessary in order to have electron transport in this configuration. The maximum value of the peak is hard to determine because it depends on the transmission coefficient, but it also depends of the overlapping between the DOS of the Qds. \\
Concerning to the array configuration, Fig.~\ref{septima}(c), the I-V curve is determined by a combination of the two previous cases. In order to have transport the resonant state condition must be fulfilled therefore a current peak appears but the total current is the sum of the partial currents of each row. 

\subsection{Comparison with NEGFF}
Finally, the results obtained using the proposed approach have been compared to the results of Sun et al. \cite{jap}. In that paper the authors have used the nonequilibrium Green's function method (NEGFF) to study the electron transport between  one, two and three Qds in several configurations. Their I-V results have been plotted in our figures (NEGFF in the legend). The main results are:
\begin{itemize}
\item The results presented  in Figs.~\ref{primera},~\ref{segunda},~\ref{quinta},~\ref{sexta}(a) and \ref{sexta}(c) are in accordance between the two approaches. For the serial configuration Fig.~\ref{segunda} the differences are due to the different values of the Qd coupling, we also obtain a resonant peak when the energy levels of the Qds are placed in a resonant state. The resonant state is strongly dependent on the capacitive coupling of the Qds, as the position of the energy level with the applied bias voltage depends on the capacitive coupling of the Qd.\\
In the parallel configuration we obtain the same staircase shape, but, in our case we also take into account the energy charge terms, therefore the steps occur at higher voltages.

\item The main difference appears in the case described in Fig.~\ref{tercera}. For this configuration Sun et al. predicts an antiresonance effect. We do not recover this effect because our model considers each Qd as a separate quantum systems. For this reason, our approach is known as a non-coherent model.

\item For the systems presented in Figs.~\ref{cuarta} and \ref{sexta}(b) we obtain similar results. The position of the current peak is different because Sun et al. assume that the bias is uniformly applied throughout the whole system meanwhile we take into account all the electrostatic coupling between the different parts of the system.

\end{itemize} 

As we have shown the electrostatic coupling plays an important role to determine the I-V characteristic of the system. The electrostatic effect has two terms: the first term is determined by the influence of the leads and the neighbor Qd and it is described by the capacitive coupling of the Qd and its surrounding. The second term takes into account the charge stored inside the Qd, this effect is related to the electron-electron interaction and the self consistent solution of \ref{pop} and \ref{pot} is the first approach to introduce many body effects, like the Coulomb Blockade. If we create nanodevices in oder to take advantage to the quantization of the current, only small number of discrete energy levels are available for conduction, the accurate control of the energy levels with the applied bias voltage is one of the most important points that we need to take into account. Therefore a good modelization of the Qd-Qd and Qd-lead capacitances is necessary. \\
This paper precedes future works in which realistic DOS, energy dependent transmission coefficients as well as a realistic capacitive couplings can be introduced. 
  
\section{Conclusion}

We propose a theoretical model to study the electron current in systems based in quantum dots (Qds). This model is based in the transfer Hamiltonian formalism and computes the I-V and N-V curves in the self consistent field regime (SCF), using non-coherent rate equations (NCRE). This approach provides a simple and transparent method to describe the electron transport. Due to the simplicity of the model, this can be easily extended to analyze arbitrary large arrays of Qds of interest in technological applications. Despite its simplicity and in contrast with other approaches the effect of self-charge has been taken into account, by solving the Poisson equation with appropriate boundary conditions for each Qd. As expected, the calculation of the local potential inside each Qd is one of the most critical points, since the I-V curves depends on the position of the energy level.\\
In order to show the potential of this method to analyze realistic configurations, we have studied the electron transport between different Qd configurations. We have also compared the NCRE results with well established data obtained with the NEGFF approach. Such a successful comparison shows that NCRE is a powerful and intuitive method to describe the electron transport.

\ack{
N. Garcia acknowledges the Spanish MICINN for her PhD grant in the FPU program. A. Cirera acknowledges support from ICREA academia program. The authors thankfully acknowledge the computer resources, technical expertise and assistance provided by the Barcelona Supercomputing Center - Centro Nacional de Supercomputaci\'on. The research leading to these results has received funding from the European Community’s Seventh Framework Programme 
(FP7/2007-2013) under grant agreement n°: 245977. }

% Create the reference section using BibTeX:
\section*{References}
\providecommand{\newblock}{}

\end{document}